\documentclass[aps,amssymb,superscriptaddress,footinbib,titlepage]{revtex4} 
\usepackage{amsmath}
\usepackage{amsfonts}
\usepackage{amsmath}
\usepackage{amsfonts}
\usepackage{graphicx}
\usepackage{epsfig}
\usepackage{subfigure}
\usepackage{bbm}
\usepackage{epstopdf}
\usepackage{psfrag}
\usepackage{pstool}
\usepackage{pdfpages}

\newcommand{\ba}{\begin{align}}
\newcommand{\ea}{\end{align}}
\newcommand{\be}{\begin{equation}}
\newcommand{\ee}{\end{equation}}
\newcommand{\bea}{\begin{eqnarray}}
\newcommand{\eea}{\end{eqnarray}}
\newcommand{\ls}{\left[}
\newcommand{\rs}{\right]}
\newcommand{\lr}{\left(}
\newcommand{\rr}{\right)}

\newcommand{\bsm}{\begin{bmatrix}}
\newcommand{\esm}{\end{bmatrix}}
\newcommand{\tr}{\text{tr}}

\newcommand{\half}{\frac{1}{2}}

\newcommand{\fmf}{\langle X_1\rangle}
\newcommand{\fms}{\langle X_2\rangle}

\DeclareMathOperator\arctanh{arctanh}
\DeclareMathOperator\arcsinh{arcsinh}

\begin{document}

\title{Motion and gravity effects in the precision of quantum clocks}
\author{Joel Lindkvist}
\email[Correspondence to: ]{joell@chalmers.se}
\affiliation{Microtechnology and Nanoscience, MC2, Chalmers University of Technology, S-41296 G\"oteborg, Sweden}
\author{Carlos Sab{\'\i}n}
\affiliation{School of Mathematical Sciences,
University of Nottingham,
University Park,
Nottingham NG7 2RD,
United Kingdom}
\author{G\"{o}ran Johansson}
\affiliation{Microtechnology and Nanoscience, MC2, Chalmers University of Technology, S-41296 G\"oteborg, Sweden}
\author{Ivette Fuentes}
\affiliation{Faculty of Physics, University of Vienna, Boltzmanngasse 5, 1090 Vienna, Austria}
\begin{abstract}
We show that motion and gravity affect the precision of quantum clocks. We consider a localised quantum field as a fundamental model of a quantum clock moving in spacetime and show that its state is modified due to changes in acceleration. By computing the quantum Fisher information we determine how relativistic motion modifies the ultimate bound in the precision of the measurement of time. While in the absence of motion the squeezed vacuum is the ideal state for time estimation, we find that it is highly sensitive to the motion-induced degradation of the quantum Fisher information. We show that coherent states are generally more resilient to this degradation and that in the case of very low initial number of photons, the optimal precision can be even increased by motion. These results can be tested with current technology by using superconducting resonators with tunable boundary conditions.
\end{abstract}
\maketitle
\section*{Introduction}
Precise time keeping is a key ingredient of countless applications in the information era, ranging from high-speed data transmission and communication to the Global Positioning System (GPS).

In the last years, quantum clocks based on optical transitions of ions or neutral atoms in optical lattices have achieved unprecedented levels of precision and accuracy \cite{wineland,ye}. Moreover, the use of entanglement will enable to overcome the current limitations and build up networks of clocks operating close to the Heisenberg limit \cite{heisenbergclocks}, the ultimate fundamental bound imposed by quantum mechanics.
\begin{figure}[t]
\centering
\includegraphics[width=\columnwidth]{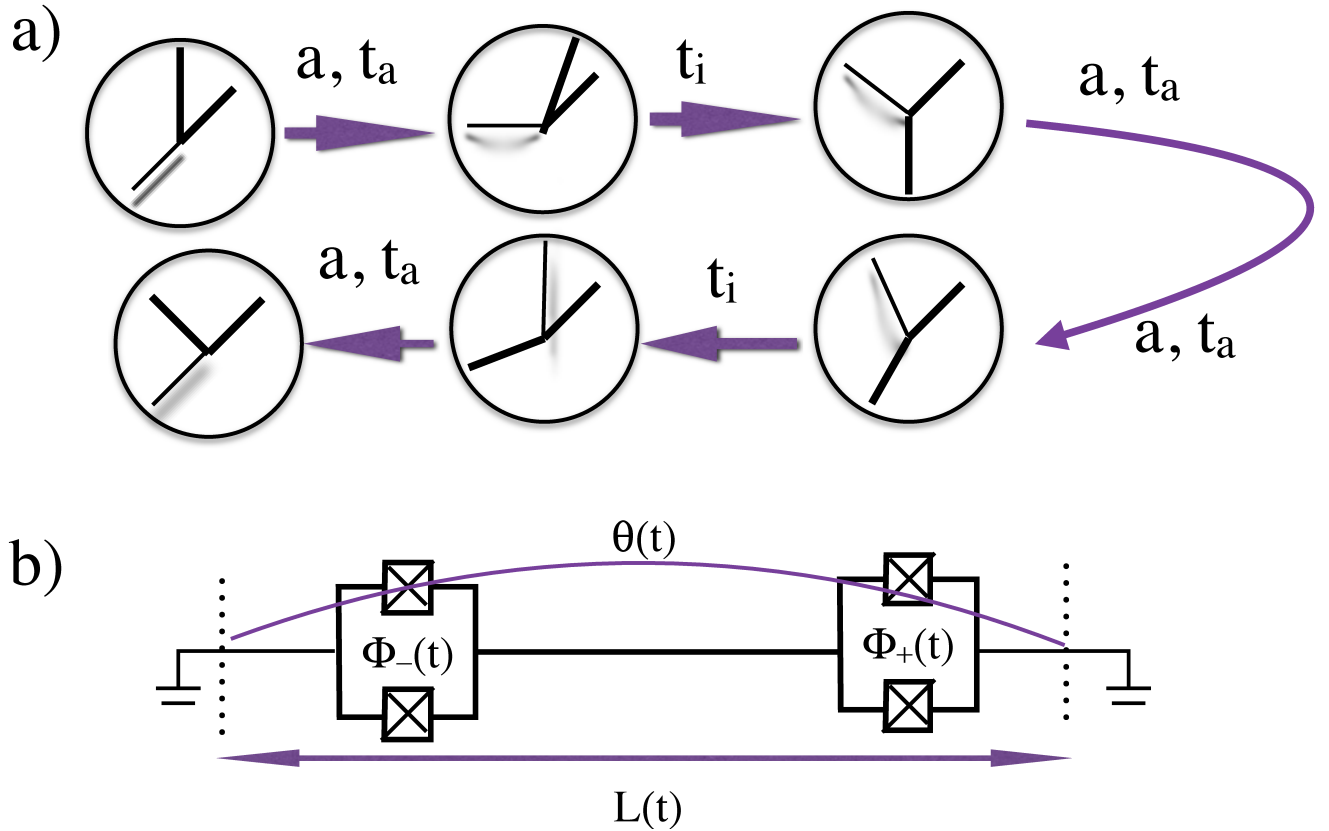}\caption[]{a) A clock undergoes a round trip characterised by four intervals of proper constant acceleration $a$ and duration $t_a$ and two intervals of constant velocity of duration $t_i$. After the trip the precision of the clock has changed. b) Experimental setup where the clock is the phase $\theta$ of a Gaussian state of the electromagnetic field in a superconducting resonator with tunable boundary conditions. A superconducting transmission line is interrupted by two SQUIDs generating a cavity of effective length $L$. The position of the effective mirrors can be moved at relativistic speeds by ultrafast variation of the magnetic fluxes $\Phi_+$, $\Phi_-$, thus the clock can undergo the trajectory depicted in a).}
\label{fig:fig0}
\end{figure}
Space agencies are planning to use ultra-precise and portable atomic clocks in space, which will allow for important new applications in fundamental physics, geophysics, astronomy and navigation. In this regime it is expected that Einstein's theory of relativity becomes relevant. This is not surprising since time dilation effects can be detected due to a difference of even less than $1\,\operatorname{m}$ in the gravitational field of the Earth \cite{chou}. However, current designs of quantum clocks are described by non-relativistic quantum mechanics.
In order to analyse the effects of gravity and motion on quantum clocks we need to work within Quantum Field Theory (QFT) since this theory allows to properly incorporate both quantum and relativistic effects.  In QFT in curved spacetime \cite{birrelldavies}, light and matter is described by quantised fields while the spacetime remains classical, which is a good approximation in the regime at which satellites operate. Tantalising predictions of QFT in curved spacetime such as Unruh-Hawking radiation or the Dynamical Casimir Effect \cite{moore} are starting to receive experimental confirmation \cite{casimirwilson}. Moreover, the use of Quantum Information and Quantum Metrology tools within a QFT framework has recently enabled the prediction of non-trivial effects of gravity, accelerated motion and spacetime dynamics on key quantum properties such as entanglement \cite{ivyalsingreview,teleportationico,fateofentanglement,entanglementincurved,localizeddetection,doukasbrown,HuLiu}. Thus it is natural to ask whether motion and gravity can affect the performance of quantum clocks. In order to address this question we need to consider a fundamental model of clock that is both quantum and relativistic, that is  a localised quantum system with periodic dynamics and whose motion through the spacetime can be properly described. Therefore, we need to consider a single mode of a localised quantum field. For the sake of simplicity we can assume that the field is confined within a cavity. The phase of this cavity mode can be used as the pointer of our clock, as we will see in more detail below.

In this paper, we show that relativistic motion affects the precision of a quantum clock. Via the equivalence principle we conclude that the same effect occurs in the case of non-uniform gravitational fields. In particular, we consider the general model of a relativistic quantum clock described above and assume that it undergoes a trajectory with non-uniform acceleration. The motion generates new particles due to the Dynamical Casimir Effect \cite{moore, casimirwilson}, together with mode-mixing among the different modes inside the cavity \cite{bsgates}. Therefore, the precision of the clock is affected. We characterise this change by computing the quantum Fisher information (QFI) of the state, which provides the fundamental bound imposed by quantum mechanics to the precision of the clock. To this end, we use the recently developed techniques of relativistic quantum metrology \cite{rqm, rqm2}. We find that while the best choice of state for estimation of time in the absence of motion is a squeezed vacuum, this state also experiences a relatively big loss of precision due to motion. Coherent states are more robust to this degradation, and in the case of very low initial number of photons we find that the precision can be even increased by motion.

The results can be readily implemented in the laboratory by using superconducting resonators with tunable boundary conditions. The boundary conditions are provided by the magnetic flux threading a SQUID, which can undergo ultrafast variations mimicking the motion of a mirror at velocities close to the speed of light, like in the first observation of the Dynamical Casimir Effect \cite{casimirwilson}. 
This setup paved the way for several tests of the interplay between quantum and relativistic effects \cite{teleportationico,twinpa}. In particular, in \cite{twinpa} we showed how to implement a test of relativistic time dilation with superconducting circuits, analysing the effects of particle creation in the twin paradox scenario. In this case, by ultrafast modulation of the electric length of the cavity, the clock experiences similar boundary conditions as in a spaceship moving at relativistic speeds. 

\section*{Cavity clock}
Let us now explain our model in more detail. As explained above, we need to consider a localised quantum field. For the sake of simplicity we will assume that the field is confined in a box-type potential. The clock will thus be a cavity containing a quantized one-dimensional electromagnetic field in a Gaussian state. The proper length $L$, i. e. length measured by a comoving observer, is constant. Although we can consider general trajectories, in order to illustrate our results we choose the trajectory so that the clock undergoes a round trip (see figure \ref{fig:fig0}a), composed of four accelerated segments and two segments of inertial motion, similar to the one of the travelling twin in the twin paradox scenario \cite{twinpa}. During each accelerated segment of duration $t_a$ in lab coordinates, an observer in the center of the cavity moves with constant proper acceleration $a$.  During the inertial segments, the observer moves with a constant velocity that is set by $a$ and $t_a$ and we denote the duration of these segments by $t_i$. Thus, the trajectory is completely described by $a$, $t_a$ and $t_i$. In the lab frame, the duration of the trip is $t_t\equiv 4t_a+2t_i$.
For an inertial observer, a 1D electromagnetic field $\phi$ obeys the Klein-Gordon equation
\be
 (\partial_t^2-c^2\partial_x^2)\phi=0,
\ee
where $c$ is the speed of light.
The two cavity mirrors introduce Dirichlet boundary conditions $\phi=0$ at two points separated by a distance $L$. Quantizing the field in Minkowski coordinates, we obtain a discrete set of cavity modes with frequencies $\omega_n=\pi n\,c/L,\hspace{3pt}n=1,2,...$.

For periods involving uniform acceleration we consider an observer moving with constant proper acceleration $a$.  This observer is static in the Rindler coordinates $(\eta,\xi)$, defined by
\bea
x&=&\frac{c^2}{a}e^{a\xi/c^2}\cosh{\lr a\eta/c\rr}\\
t&=&\frac{c}{a}e^{a\xi/c^2}\sinh{\lr a\eta/c\rr}.
\eea
In these coordinates, the wave equation is also a Klein-Gordon equation. Thus the quantization of the field gives rise to a similar set of cavity modes. The mirrors introduce Dirichlet boundary conditions at two points separated by a distance $L'=\frac{c^2}{a}\arctanh{\lr \frac{aL}{2c^2}\rr}$ with respect to Rindler position $\xi$, corresponding to a proper distance $L$, and the mode frequencies with respect to Rindler time $\eta$ are $\Omega_n=\pi n\,c/L',\hspace{3pt}n=1,2,...$.

The initial and final states of the cavity are related by a Bogoliubov transformation which in this case is a combination of the Bogoliubov transformation between the inertial and uniformly accelerated modes and the phases acquired during the free evolution \cite{bruschi}. More specifically, before the trip, the modes in the cavity are described by a set of annilhilation and creation operators, $a_n$ and $a_n^{\dagger}$, satisfying the canonical commutation relations $[a_m,a_n^{\dagger}]=\delta_{mn}$. The modes in the cavity after the trip are similarly described by another set of operators, $b_n$ and $b_n^{\dagger}$, satisfying similar commutation relations. These two sets are related by a Bogoliubov transformation, defined by
\be
b_m=\sum_n\lr A_{mn}^*a_n-B_{mn}^*a_n^{\dagger}\rr.\label{eq:bogodef}
\ee
The Bogoliubov coefficients $A_{mn}$ and $B_{mn}$ are functions of the trajectory parameters $a$, $t_a$ and $t_i$ and the proper length $L$ of the cavity. They can be computed analytically as power series expansions in the dimensionless parameter $h\equiv aL/c^2$ \cite{twinpa}.

The first mode of the cavity is prepared in a Gaussian state, with vacuum in the higher modes. Free time-evolution of a Gaussian state corresponds to a phase rotation. Since the proper length of the cavity is preserved throughout the trip, that is true also for the mode frequencies. Thus, we can relate the accumulated phase shift during the trip to an elapsed proper time by simply dividing with the frequency of the first mode. This allows us to use the phase of the single-mode state of the electromagnetic field in the cavity as a clock.

\section*{Clock precision}
Now, let us introduce quantum metrology tools for the computation of the optimal bounds to the precision of the clock. We will take advantage of the elegance of the covariance matrix formalism for Gaussian states. Initially preparing the cavity in a Gaussian state ensures that the final state will also be Gaussian.  A Gaussian state is completely characterized by the first moments, $\langle X_n\rangle$, and the covariance matrix
\be
\sigma_{mn}=\half\langle \{X_m,X_n\}\rangle-\langle X_m\rangle\langle X_n\rangle.
\ee
We define the quadrature operators by $X_{2n-1}=\frac{1}{2}(a_n+a_n^{\dagger})$ and $X_{2n}=-\frac{i}{2}(a_n-a_n^{\dagger})$. In the following, we will consider only the case where we start with the cavity in a single-mode state, since we can use the phase of one mode with fixed frequency as the pointer of our clock. We will focus only on that particular mode after the transformation, and trace out the other modes. As stated above, the fundamental mode of the cavity is used as the clock mode.
A single-mode Gaussian state can always be parametrized by the real displacement parameter $\alpha$, the complex squeezing parameter $\xi=re^{i\phi}$ and the phase $\theta$, as well as the purity $P$. The position in phase space is encoded in the first moments
\bea
\alpha&=&\sqrt{\langle X_1\rangle^2+\langle X_2\rangle^2}\\
\tan{\theta}&=&\langle X_2\rangle/\langle X_1\rangle, 
\eea
while the squeezing $\xi$ and the purity $P$ are encoded in the covariance matrix $\sigma$:
\bea
P&=&\frac{1}{4\sqrt{\det{\sigma}}}\nonumber\\
r&=&\half\arctanh{\lr \frac{\sqrt{(\sigma_{11}-\sigma_{22})^2+(2\sigma_{12})^2}}{\sigma_{11}+\sigma_{22}}\rr}\nonumber\\
\tan{\lr2\theta+\phi\rr}&=&\frac{2\sigma_{12}}{\sigma_{11}-\sigma_{22}}.
\eea

The QFI of a state quantifies the maximum precision that can be achieved in the estimation of a parameter encoded in the state. The QFI for estimation of a parameter $\tau$ using a single-mode Gaussian state is given by \cite{braun}
\bea
H_{\tau}&=&X'^{T}(\tau)\sigma^{-1}(\tau)X'(\tau)+\frac{1}{2}\frac{\tr{\ls\lr\sigma^{-1}(\tau)\sigma'(\tau)\rr^2\rs}}{1+P(\tau)^2}
+2\frac{P'(\tau)^2}{1-P(\tau)^4},
\label{Fishergeneral}
\eea
where the prime denotes a derivative with respect to $\tau$. Now, as explained above, our parameter of interest is the phase $\theta$, since we will use it as the pointer of our clock.
Expressed in the state parameters, the QFI is
\be
H_{\theta}=4\alpha^2P\ls\cosh{\lr2r\rr}+\sinh{\lr2r\rr}\cos{\phi}\rs+\frac{4\sinh^2{\lr 2r\rr}}{1+P^2}.\label{Fisherparameters}
\ee
We note that this expression is independent of $\theta$.
The mean number of photons for a general single-mode Gaussian state can be written as 
\be
N=\alpha^2+\frac{1}{2}\lr\frac{1}{P}-1\rr+\frac{\sinh^2{\lr r\rr}}{P}.
\ee
 Optimizing (\ref{Fisherparameters}) with a fixed photon number, it can be shown that the choice of state that maximizes the quantum Fisher information is the squeezed vacuum \cite{braun}. In other words, all available energy should be put into squeezing for the best possible phase estimation.

The Cram\'er-Rao inequality gives a lower bound on the mean-square error of parameter estimation and is satisfied for optimal measurements. For $M$ optimal measurements, this yields the following expression for the phase variance
\be
\Delta\theta=\frac{1}{\sqrt{MH_{\theta}}}.\label{cramer}
\ee
 
\section*{Bogoliubov transformation}
Let us now examine how the QFI of an initial single-mode Gaussian state is affected by the motion described by a Bogoliubov transformation with coefficients $A_{mn}$ and $B_{mn}$. 
The reduced covariance matrix of mode $k$ after the Bogoliubov transformation can be written as \cite{rqm2}
\be
\sigma_{k}(\theta)=\mathcal{M}_{kk}(\theta) \sigma_{0} \mathcal{M}_{kk}^{T}(\theta)+\frac{1}{4}\sum_{n\neq k}\mathcal{M}_{kn}(\theta)\mathcal{M}_{kn}^{T}(\theta),\label{eq:covmatrixtransf}
\ee
where $\sigma_0$ is the initial covariance matrix and $\mathcal{M}_{mn}$ are the $2\times2$ matrices
\be
\mathcal{M}_{mn}=\left(
                   \begin{array}{cc}
                     \Re(A_{mn}-B_{mn}) & \Im(A_{mn}+B_{mn}) \\
                     -\Im(A_{mn}-B_{mn}) & \Re(A_{mn}+B_{mn})
                   \end{array}
                 \right)\,.
\ee
 Here $\Re$ and $\Im$ denote the real and imaginary parts, respectively. The first moments transform as
\bea
\fmf&=&Re\lr A_{11}-B_{11}\rr\fmf_0+Im\lr A_{11}+B_{11}\rr\fms_0\nonumber\\
\fms&=&-Im\lr A_{11}-B_{11}\rr\fmf_0+Re\lr A_{11}+B_{11}\rr\fms_0\nonumber.\\
\label{eq:firstmomentstransf}
\eea
The main aim of this paper is to compare the QFI of the state after the motion, described by the covariance matrix (\ref{eq:covmatrixtransf}) and the first moments (\ref{eq:firstmomentstransf}),  with the one of the initial state described by $\sigma_0$, $\fmf_0$ and $\fms_0$. This allows us to analyse the effects of relativistic motion in the precision of a clock. 
To qualitatively describe the effects, it will be enough to consider the so-called building block transformation. This transformation consists of the Bogoliubov transformation between inertial and accelerated modes, a phase shift $k\theta_a=k\pi\arcsinh{\lr at_a/c\rr}/(2\arctanh{\lr h/2\rr})$ of mode $k$ acquired during the acceleration, and the inverse of the first transformation. For details on how to compute the coefficients, see \cite{twinpa}. The transformation, and thus all the quantities of interest, are $2\pi$-periodic in $\theta_a$. The transformation for a more general trajectory can be composed of such building-block transformations, with phase shifts $k\theta_i$ for the inertial motion in between. 

As discussed in \cite{twinpa}, the rate of the cavity clock is modified by uniform acceleration. This is due to the fact that different points in the cavity experience different proper times, and the effect can be understood classically. On top of that, however, there is also an extra phase shift due to mode-mixing and particle creation that depends on changes in acceleration.
Since the QFI in equation (\ref{Fisherparameters}) is independent of the phase, it stays constant during free Minkowski or Rindler time-evolution. Thus, it is only affected by the mode-mixing and particle creation induced by changes in acceleration.

\section*{Results}
\begin{figure}[ht]
\centering
\includegraphics[width=\columnwidth]{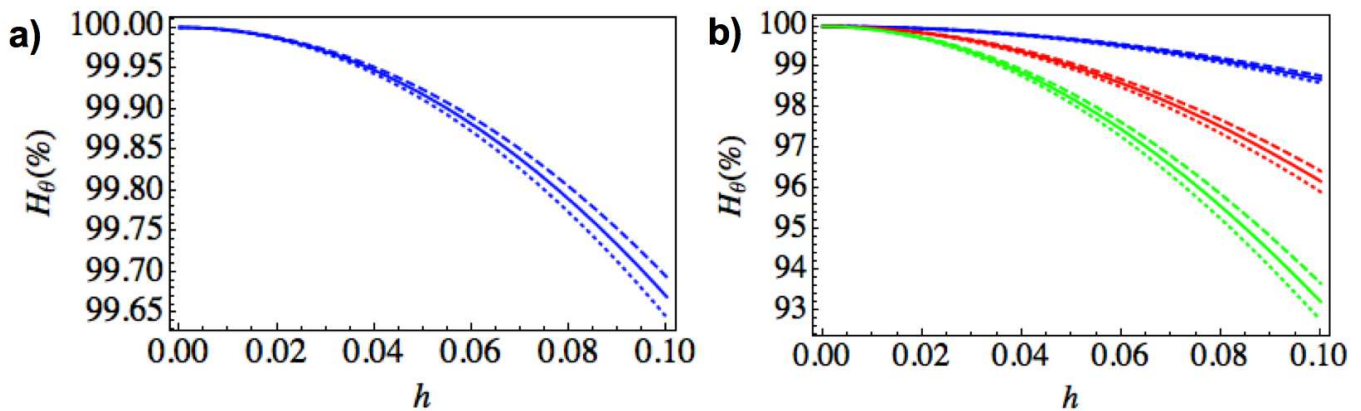}\caption[]{Ratio of the transformed and original QFI as a function of $h$, for $\theta_a=\pi$. In a), the initial state is coherent with $N>1$. In b), the initial state is the squeezed vacuum with $N=1$ (blue), $N=5$ (red) and $N=10$ (green). The solid curves show the effect of mode-mixing, while the dotted (dashed) curves include the effects of particle creation for $\theta_0=0(\pi/2)$.}
\label{fig:fig2}
\end{figure}

\begin{figure*}[ht]
\centering
\includegraphics[width=\columnwidth]{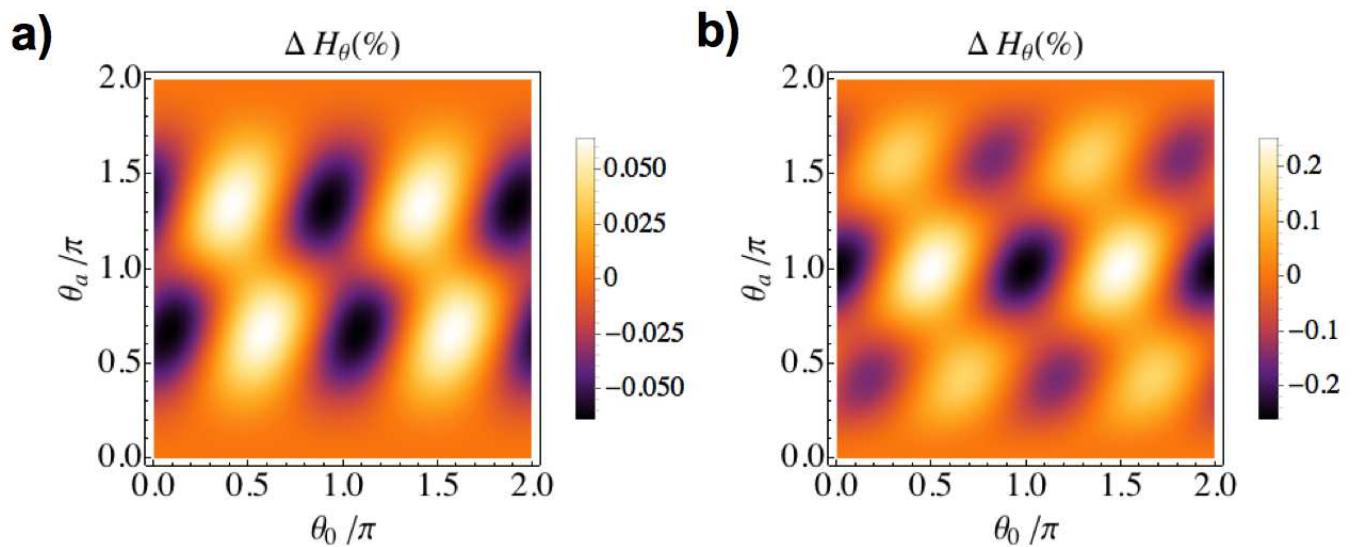}\caption[]{Difference in transformed QFI with and without particle creation coefficients, as a function of the transformation angle $\theta_a$ and initial phase $\theta_0$ and normalized to the initial QFI. In a), the initial state is coherent with $N>1$. In b), the initial state is the squeezed vacuum with $N=5$. The transformation parameter is $h=0.1$.}
\label{fig:fig25}
\end{figure*}

We start by considering the QFI in the case when the initial state is coherent, with a mean photon number $N=\alpha_0^2$. In figure \ref{fig:fig2}a we plot the ratio of the QFI for a clock having undergone motion and an inertial clock. The motion of the cavity generates mode-mixing between the clock mode and the higher modes. Tracing out these consequently leads to a suppression of the displacement parameter $\alpha$ and the purity $P$, resulting in the degradation of the QFI seen in figure \ref{fig:fig2}a. This effect is independent of the initial phase $\theta_0$ and largest for $\theta_a=\pi$. Apart from mode-mixing, particle creation effects also lead to a shift in the QFI, which may be positive or negative depending on $\theta_0$ and $\theta_a$. This is a genuine quantum effect affecting the precision of the clock. To estimate the magnitude of the effect, we compare the QFI with the particle creation coefficients in the Bogoliubov transformation neglected, to the QFI obtained using the full transformation. In figure \ref{fig:fig25}a we plot this difference as a function of $\theta_0$ and $\theta_a$. In the regime interesting for clock purposes ($N>1$), the QFI degradation of a coherent state is independent of the initial mean photon number.

Next, we treat the case when the initial state is the squeezed vacuum, with a mean photon number $N=\sinh^2{r_0}$. Now, only the second term in equation (\ref{Fisherparameters}) is relevant. In figure \ref{fig:fig2}b, we plot again the ratio of the transformed and initial QFI. Here, the mode-mixing leads to a suppression of the squeezing parameter $r$ and a corresponding QFI degradation. Figure \ref{fig:fig25}b shows the shift due to particle creation, computed in the same way as for the coherent state.

The QFI degradation for the two classes of initial states above show similar traits. The main difference is that, for the squeezed vacuum, it depends on the initial mean photon number $N$ and is generally larger than for the coherent state in the regime of interest. The reason for this is that the QFI scales differently with $N$ in the two cases. An equal percentage of the clock mode photons is lost due to mode-mixing, independently of $N$ and the type of state. In the coherent case, the QFI is proportional to $N$, while for the squeezed vacuum it is not.

\begin{figure}[ht]
\centering
\includegraphics[width=0.92\columnwidth]{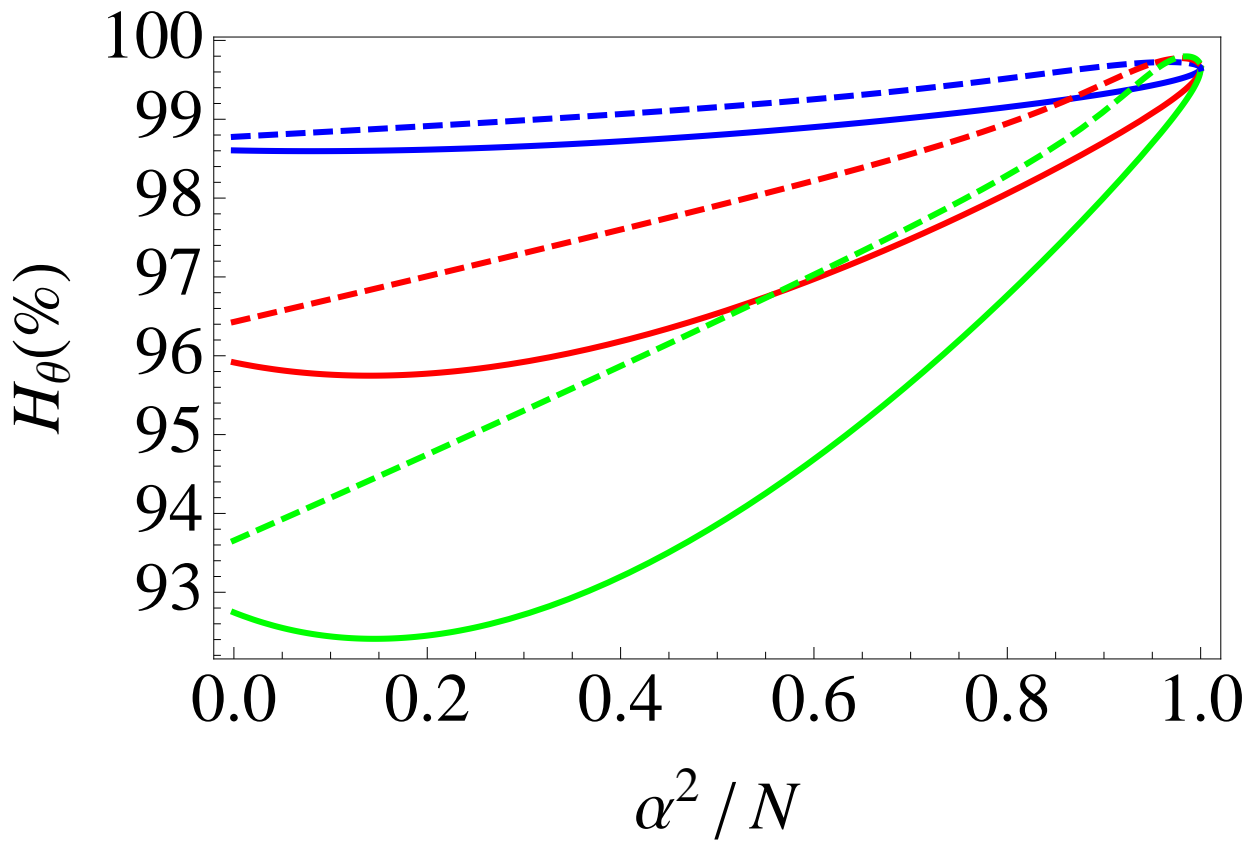}\caption[]{Ratio of the transformed and original QFI as a function of how the photons are distributed between displacement and squeezing. $\alpha_0^2/N=1$ corresponds to a coherent state and $\alpha^2_0/N=0$ to the squeezed vacuum. The transformation parameters are $h=0.1$ and $\theta_a=\pi$. The initial number of photons is $N=1$ (blue), $N=5$ (red) and $N=10$ (green). The solid(dashed) curves are for $\phi_0=0(\pi)$}
\label{fig:fig3}
\end{figure}

Let us now consider more general initial states, containing both displacement and squeezing, while we keep the mean photon number $N=\alpha_0^2+\sinh^2{r_0}$ constant. Figure \ref{fig:fig3} shows the QFI ratio for different mean photon numbers in the initial state, as a function of how the photons are distributed between displacement and squeezing. In the intermediate cases, we see an interplay between the two terms in equation (\ref{Fisherparameters}). There are qualitative differences between the initially phase squeezed ($\phi_0=0$) and amplitude squeezed ($\phi_0=\pi$) states. In the case of phase(amplitude) squeezing, the first term in equation (\ref{Fisherparameters}) increases(decreases) with the squeezing, leading to the local minima(maxima) in figure \ref{fig:fig3}. In general, the degradation of the QFI tends to be larger for squeezing-dominated states. As mentioned before, however, the squeezed vacuum is the optimal single-mode clock state. In the regime considered here, it would still be the best choice, despite the increased degradation.

\begin{figure}[ht]
\centering
\includegraphics[width=\columnwidth]{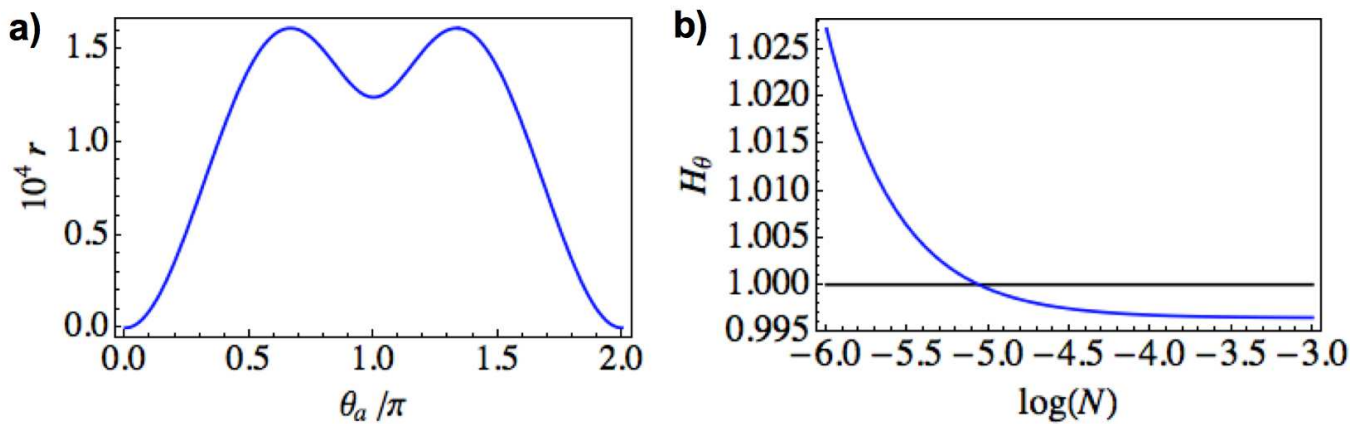}\caption[]{a) Squeezing $r$ as a function of $\theta_a$ with vacuum as the initial state, for $h=0.1$. b) Ratio of the transformed and original QFI for an initial low-power coherent state as a function of $N$, for $h=0.1$ and $\theta_a=\pi$.}
\label{fig:fig4}
\end{figure}

So far, we have only discussed initial states with $N>1$ since these are the ones interesting for clock purposes. The QFI, and thus the clock precision, generally increases with $N$. The main effect of the motion is that the clock mode loses photons due to mode-mixing, resulting in a degraded QFI. In the case of smaller $N$, however, the particle creation effects are more dominant. Starting with the vacuum ($N=0$), the trip generates a certain amount of squeezing (see figure \ref{fig:fig4}a) and an associated QFI. For small enough initial $N$, this is enough to enhance the QFI (see figure \ref{fig:fig4}b). 
In the regime treated in this paper ($h\leq 0.1$), these effects are seen only for very small $N$. For larger $h$, however, we expect to see this kind of effects also for higher-power states. In order to examine the case of larger $h$, though, we need to abandon our perturbative treatment of the Bogoliubov coefficients, which is out of scope for this paper.

\section*{Photon leakage}
In this section, we analyse how the precision of the moving clock is affected in the recent proposal of an experimental test of the twin paradox with circuit QED \cite{twinpa}. In \cite{twinpa} the accelerated clock is implemented by a superconducting resonator consisting of a transmission line interrupted by two SQUIDs (see figure \ref{fig:fig0}b), which provide highly tunable boundary conditions that can be used to let the clock experience the changes in boundary condition of a round-trip trajectory. As concluded in \cite{doukaslouko}, the correspondence between the SQUID boundary condition and a moving mirror is valid only when the effective length modulation is small compared to the wavelength of the field. In this round trip scenario, the circuit parameters can be chosen so that we safely stay within that regime. Moreover, in this more realistic scenario, photon leakage from the cavity will also degrade the QFI. For the experimental regimes suggested in \cite{twinpa}, corresponding to a maximal $h$-value of $7\times 10^{-3}$, the QFI degradation effects discussed in the previous section are small compared to the effect of photon leakage.

Let us consider that before the trip, the cavity is prepared in a state with a mean photon number $N_{i}$. For a trip of time $t_{tot}$, the number of photons at the end would be approximately 
$N_f=N_ie^{-t_{tot}/\tau}$, where the decay time $\tau$ can be written as
$
\tau=\frac{2 Q}{\omega}$
and $Q$ is the Q-value of the cavity.
In order to treat the SQUID as a tunable boundary condition, we need to stay in the regime $\phi<<2\pi$, where $\phi$ is the phase drop over the SQUID. By the Josephson relation $I=I_c\sin{\phi}$, this sets a limit on the ratio of the current $I$ through the SQUID and the effective flux-dependent critical current, given by
$I_c(\Phi_{ext})=2I_c\bigg|\cos{\lr\pi\frac{\Phi_{ext}}{\Phi_0}\rr}\bigg|,
$
where $I_c$ is the critical current of each Josephson junction in the SQUID. Since the current through the SQUID depends on the number of photons in the cavity, this also limits the maximal photon number and thus the precision in estimating the phase.
Let us now estimate the number of photons $N$ corresponding to a current $I$. The effective inductance for each SQUID is
$
L_{SQ}=\frac{\Phi_0}{2\pi}\frac{1}{I_c(\Phi_{ext})}\frac{1}{\cos{\phi}}
$
and for the cavity $L_{cav}=L_0\lambda$, where $L_0$ is the inductance per unit length and $\lambda$ the physical cavity length, related to $L$ by taking into account the initial external fluxes through the SQUIDs. Using the standard expression for the energy stored in an inductor, the number of photons in the cavity can thus be expressed as
\bea
N&=&\frac{1}{2\hbar\omega}\ls L_{cav}+2L_{SQ}\rs I^2\nonumber\\
&=&\frac{1}{2\hbar\omega}\ls L_0\lambda+\frac{\Phi_0}{\pi}\frac{1}{I_c(\Phi_{ext})}\frac{1}{\sqrt{1-(I/I_c(\Phi_{ext}))^2}}\rs I^2.\nonumber\\
\eea
Now, setting the current to $I=\kappa I_c(\Phi_{ext})$, $\kappa<1$, 
we observe that this function increases with $I_c(\Phi_{ext})$. In order to determine the maximal number of photons that can be stored in the cavity, we should thus use the smallest value of $I_c(\Phi_{ext})$, or equivalently the largest value of the external flux $\Phi_{ext}$.

\begin{figure}[ht]
\centering
\includegraphics[width=\columnwidth]{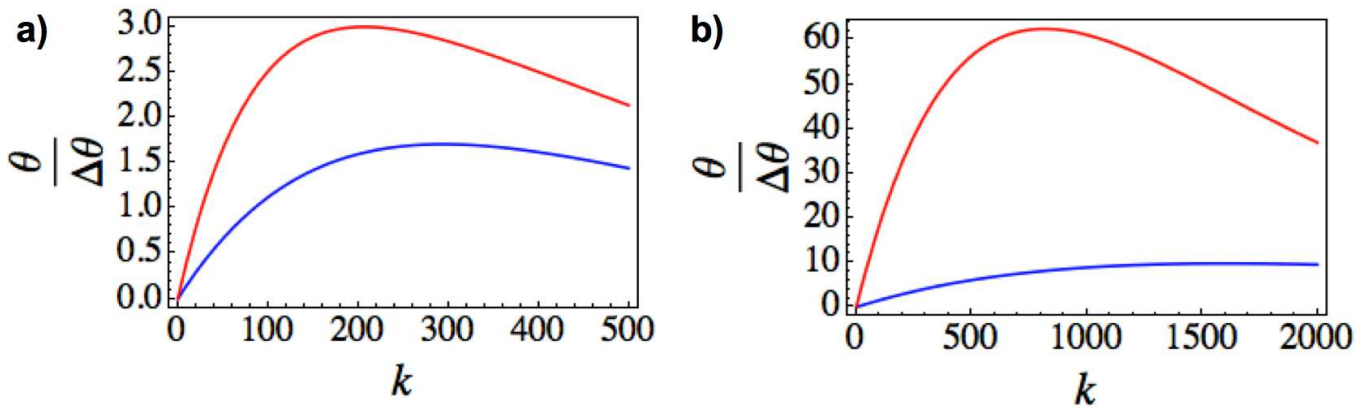}\caption[]{Signal-to-noise ratio $\theta/\Delta\theta$ for phase estimation as a function of the number of trajectories $k$. The plot in a) (b) is for the case of $L=1.1(6)$ cm and a corresponding maximal initial photon number of $N=2.98(78.5)$. The blue(red) curves are for a coherent state(squeezed vacuum).}
\label{fig:fig6}
\end{figure}

In \cite{twinpa} we considered two different scenarios, corresponding to $L=1.1$ cm and $L=6$ cm. Now, for $L_0=0.44\mu$F/m, $I_c=0.5\mu$A, $\Phi_{ext}=0.4\phi_0$ and $\kappa=0.2$, we obtain a maximal photon number of $N=2.98(78.5)$ in the short(long) cavity case.
The number of photons available at the measurement stage depends on the total travel time. For the experimental values suggested in \cite{twinpa}, each trajectory lasts $4$ ns and leads to a phase shift of $4.55\times10^{-3}(0.94\times 10^{-3})$ for the short(long) cavity. Increasing the number of trajectories leads to a larger phase shift, but also to a worse precision by decreasing the number of available photons. By using (\ref{cramer}) with $M=1$ we can compute the phase variance $\Delta\theta$ for one optimal measurement. In figure \ref{fig:fig6}, we plot the ratio of the phase shift and the phase variance (signal-to-noise ratio) as a function of the number of trajectories $k$. We assume a cavity Q-value of $10000$. For a given state, there is a certain value of $k$ maximizing the signal-to-noise ratio.

\section*{Summary and conclusions}
In summary, we show that motion and gravity can modify the fundamental bounds imposed by quantum mechanics in the measurement of time. We compute the QFI of the state of the electromagnetic field in a cavity that undergoes non-uniform accelerated motion for several initial Gaussian states. While squeezed vacuum is the optimal state for time estimation in the absence of motion, we find that it is also relatively sensitive to the loss of precision induced by motion. Coherent states are more robust to this effect, and we find that for a very low number of initial photons the QFI is even increased with motion. Our results can be tested with current technology by using superconducting resonators with tunable boundary conditions. This low-cost Earth-based experiment will inform the ongoing projects involving space-based ultra-precise quantum clocks. Moreover, we show that the application of quantum metrology tools to QFT allows us to deepen our understanding on the fundamental limits imposed by quantum mechanics in the measurement of spacetime parameters.

\section*{Acknowledgments}
We thank I-.M. Svensson for valuable discussions.  J. L. and G. J. would like to acknowledge funding from the Swedish Research Council and from the EU through the ERC. I. F. and C.S acknowledges support from EPSRC (CAF Grant No. EP/G00496X/2 to I. F.).

\section*{Author contributions}
IF and GJ conceived the project. JL performed most of the computations. CS and GJ provided theoretical assistance. IF supervised the project. All the authors contributed to the preparation of the manuscript.

\section*{Competeing financial interests statement}
The authors declare no competing financial interests.

\end{document}